\documentclass[letterpaper,11pt]{JHEP}
\usepackage{epsfig}
\usepackage{epsfig}
\usepackage{graphicx}

\preprint{{\tt hep-th/0701108}}

\title{Three dimensional Janus and time-dependent black holes}

\author{Dongsu Bak\\
        Department of Physics, University of Seoul,
        Seoul 130-743, Korea\\
        Email: \email{dsbak@mach.uos.ac.kr }}
\author{Michael Gutperle\\
        Department of Physics and Astronomy, UCLA,
 Los Angeles, CA 90055, USA\\
        Email: \email{gutperle@ucla.edu}}
\author{Shinji Hirano\\
        The Niels Bohr Institute,
Blegdamsvej 17, DK-2100 Copenhagen, Denmark\\
        Email: \email{hirano@nbi.dk}}

\abstract{
We show that the three dimensional Janus geometry can be embedded
into the type IIB supergravity and discuss its dual CFT description.
We also find exact solutions of time dependent black holes with a nontrivial
dilaton field in three and higher dimensions as an application of the
 Janus construction.}
{}
\begin{document}
\baselineskip16pt
\def\nn{\nonumber}
\section{Introduction}

It is of great interest to study nonsupersymmetric deformations of
 $AdS_5\times S^5$ space. The ultimate goal would be to find the
 gravity duals of ${\cal N}=0$ Yang-Mills theory and realistic QCD models.
The Janus deformation of $AdS_5$ space \cite{Hirano} is
nonsupersymmetric. However, its dual field theory is not in the
universality class of confining gauge theories. So in that sense
it does not meet the goal that one might hope for.
Nevertheless it is an interesting and rare example of nonsupersymmetric
deformations where both gravitational and dual field
theory descriptions are under good control.
Indeed both descriptions are remarkably simple. In the gravity side
the Janus deformation is a thick $AdS_4$-sliced domain wall in
$AdS_5$ with the varying dilaton, where asymptotically the dilaton
approaches a constant in one half of the boundary space and the
different value in the other half.
In the gauge theory side ${\cal N}=4$ super Yang-Mills (SYM)
theory is deformed by the exactly marginal operator dual to the
 dilaton -- the SYM Lagrangian up to the total derivative -- with a
space dependent deformation parameter. In effect the coupling
constant jumps discontinuously at the interface of two halves
 of the boundary space.
By construction the Janus deformation preserves the $SO(3,2)$
 symmetry of the $AdS_4$ slices. Correspondingly the dual field theory
preserves the conformal symmetry at the interface, defining the
interface conformal field theory (ICFT).
Albeit being nonsupersymmetric, the Janus deformation was shown
to be stable against a large class of perturbations and believed
to be nonperturbatively stable, owing to the existence of formal
killing spinors \cite{Freedman}.
Somewhat surprisingly, an exact agreement was found even at the
more quantitative level:
The Janus deformation predicts the vev of the exactly marginal
operator at large $N$ and large 't Hooft coupling. Meanwhile the
dual ICFT allows us to compute the vev in all orders in 't Hooft
 coupling by using the conformal perturbation theory. In fact two
results agree at the leading order in the deformation
parameter \cite{Karch}.
It is remarkable to find an exact agreement between two sides
of the strong-weak coupling duality in the nonsupersymmetric theory.

We have been stressing the nonsupersymmetric nature of Janus
and its tractability nonetheless.
It is, however, worth mentioning that the supersymmetric Janus
deformation was found in 5-dimensional gauged
supergravity \cite{Clark} and more recently in the full type
IIB supergravity \cite{D'Hoker}.
Correspondingly the ICFT can also be made supersymmetric by
introducing the interface interactions \cite{Karch}. In fact
all the possible interface interactions which yield the
supersymmetric ICFTs were classified in \cite{D'Hoker1}.
The supersymmetric Janus is interesting on its own. In particular
 it suggests the interpretation of the SUSY Janus in terms of
 the intersecting D3 and D5-branes -- a potential new decoupling
limit of intersecting D-branes. This may be of relevance in
connection to the D-brane realization of Karch-Randall model
\cite{KarchR}.

The nonsupersymmetric Janus allows several generalizations.
 The Janus type domain wall exists in arbitrary dimensions
and exhibits pseudo-supersymmetries
\cite{Freedman, Behrndt:2002ee, Celi, Skenderis:2006jq}. In the
type IIB case the axion can be turned on by the $SL(2,Z)$
rotation \cite{D'Hoker}. The $AdS_d$-sliced Janus can accommodate
the $AdS_{p < d}$-sliced Janus(es) within it in a self-similar
 fashion \cite{Hirano1}. By a double analytic continuation the
Janus geometry can be utilized to argue a dual of the Big
Bang/Crunch cosmology from the AdS/CFT perspective \cite{Bak:2006nh}.

In this note we wish to extend our previous study of the
Janus deformation to the $AdS_3\times S^3\times M_4$ space.
 The $AdS_2$-sliced Janus was previously discussed
in \cite{Freedman}.
Our aim is to embed it into the AdS/CFT setup in the 10d type
IIB string theory. This will thus yield the Janus deformation
of the AdS$_3$/CFT$_2$ correspondence -- Janus$_3$/ICFT$_2$.
It is our hope that the further simplicity due to the low
dimensionality facilitates more quantitative studies and
provides new qualitative perspectives.

Besides the Janus deformation of the
$AdS_3\times S^3\times M_4$ space,
we also discuss the application of the $AdS_2$-sliced Janus
to the black hole. Exploiting the fact that the BTZ black hole
is a quotient of the $AdS_3$ space, it is rather straightforward
to consider the Janus deformation of the BTZ black hole.
We will see that the Janus BTZ black hole is time
dependent and has two disconnected boundaries in which the dilaton
takes different constant values. We generalize this
by constructing
the higher dimensional black holes.  The three and five dimensional
solutions can be embedded into the 10d type IIB supergravity.

The paper is organized as follows.
In section 2 we construct the Janus
deformation of the $AdS_3\times S^3\times M_4$ space.
In section 3 we briefly discuss its dual CFT interpretation.
In section 4 we discuss the Janus deformation of the BTZ black hole.
In section 5, we deal with the higher dimensional generalization of the
time dependent black hole solution.
We conclude our discussions in last section.

\section{Janus deformation in three dimensions }

In this section we would like to discuss the Janus
deformation of the AdS$_3 \times S^3\times M_4$, where $M_4$
may be taken as either $T^4$ or $K3$. As we shall see below
the deformation along
the internal  $M_4$ directions will be just a warping by a
conformal factor related to the dilaton. Thus the details of the
internal geometry do not play any role in this study.
The spirit of writing down the ansatz will be
pretty much the same as the case of AdS$_5\times S^5$. Along
 the deformation, we like to
keep the  $SO(1,2)\times SO(4)$ part out of the original
$SO(2,2)\times SO(4)$ global symmetries.  One complication
is that there is a possibility of adding an extra warp
factor along
the internal dimensions. However,
it turns out that
the warp factor does not play a role of an
extra degree of freedom. 
Rather it is
determined uniquely as a function of dilaton
by imposing the $SO(1,2)\times SO(4)$ part
of the global symmetries.

We take the ansatz for the Janus solution in the Einstein frame  given by
\begin{eqnarray}
ds^2 &=& e^{\phi\over 2}f(\mu) \left(d\mu^2 + ds^2_{AdS_{2}}\right)+
      e^{\phi\over 2}\,ds_{S^3}^2 + e^{-{\phi\over 2}}
\,ds^2_4
\ ,
\nn\\
\phi&=&\phi(\mu)\ , \label{ansatza}\\
F_3&=& 2 f(\mu)^{3\over 2} \, d\mu \wedge \omega_{AdS_2}
+ 2 \omega_{S^3}\ ,\nn
\end{eqnarray}
where $\omega_{AdS_2}$ and $\omega_{S^3}$ are the unit volume forms
on $AdS_2$ and $S^3$ respectively.
The line element $ds_4^2$ is
for the internal manifold $M_4$, which may be
either $T^4$ or $K3$.


 The relevant IIB supergravity equations of motion are
given by
\begin{eqnarray}
&&R_{\alpha\beta} -{1\over 2}\partial_\alpha \phi
\partial_\beta \phi
  -{1\over 4} e^\phi F_{\alpha}^{\phantom{a}\mu\nu}
F_{\beta\,\mu\nu}+{1\over 48} e^\phi F^2 g_{\alpha\beta}
=0\ ,\nn\\
&& \nabla^2 \phi = {1\over 12} e^\phi F^2
\ ,\label{eqofm}\\
&& \nabla_\alpha (e^\phi F^{\alpha\beta\gamma} )=0\ ,\nn
\end{eqnarray}
which should be supplemented by the Bianchi identity $dF_3=0$.
The equation of motion for the dilaton can be integrated
leading to
\begin{equation}
\phi'(\mu) = {\gamma\over f^{1\over 2}(\mu)}\ .
\label{dila}
\end{equation}
The Einstein equations give rise to
\begin{eqnarray}
 f'f' - f f'' &=& -2 f^3 +\gamma^2 
f
 \ ,\nn\\
f'f' - 2 f f''&=& -8  f^3+4 f^2\,.
\label{einstein}
\end{eqnarray}
It is easy to see that these equations are equivalent to the first
order differential equation
\begin{equation}
f'f' = 4 f^3 -4 f^2+ 2\gamma^2 f\ ,
\label{einsteinb}
\end{equation}
corresponding to the motion of a particle with zero  energy in a
potential given by 
\begin{equation}
V(f) = -4 f \left(f^2 -f+{\gamma^2 \over 2}\right)\ .
\label{potential}
\end{equation}

So far we have been working with the 10d equations of motion
but the above final equation
may also  be derived from  a dimensionally  reduced action
down to three dimensions.
We take the ansatz for the dimensional reduction as
\begin{eqnarray}
ds^2 &=& e^{\phi\over 2} g_{ab}dx^a dx^b
      +
      e^{\phi\over 2}\,ds_{S^3}^2 + e^{-{\phi\over 2}}
\,ds^2_4
\ ,
\nn\\
\phi&=&\phi(x)\ , \label{reduction}\\
F_3&=& 2 (\omega_{g_{ab}}
+ 
\omega_{S^3})\ ,\nn
\end{eqnarray}
where
 we denote the three metric and its volume form by
$ds_3^2=g_{ab}dx^a dx^b$ and  $\omega_{g_{ab}}$ respectively.
The three metric and  the dilaton can  be a general function of
the three coordinates $x^a$. Upon the dimensional reduction,
the IIB supergravity becomes
the Einstein gravity coupled to a scalar
with a negative
cosmological constant; the resulting  action reads 
\begin{equation}
I=  {1\over 16\pi G_3}\int d^3x \sqrt{g_3}\left(R_3- 
g_3^{ab}\partial_a\phi
\partial_b\phi + 2
\right)\ .
\label{3action}
\end{equation}
where $G_3$ is the 3d  Newton constant. 
We follow
the convention of \cite{deBoer}, where the 3d AdS radius and the 3d
Newton constant  are related to the D1/D5 charges, $Q_1$/$Q_5$, by
\begin{equation}
R^2_{ads}= g_6 \sqrt{Q_1 Q_5}\, l_s, \ \ \ \ \ \ \
G_3= { \sqrt{g_6}\over 4(Q_1Q_5)^{3/4} }  \, l_s\,,
\end{equation}
where  the six dimensional
string coupling $g_6$ is related to the 10d string coupling by
$g_6^2= g^2 Q_5/Q_1$. The supergravity description is valid if
$g_6 Q_1$ and $g_6 Q_5$ are large but fixed.
We shall  set $R_{ads}=1$.

Let us solve our main equation (\ref{einsteinb}).
If $\gamma^2 > 1/2$, the geometry develops a  naked
curvature singularity.
We shall restrict below our discussion to
the  case of $\gamma^2 < 1/2 $ unless otherwise is mentioned 
specifically.
The roots of the polynomial,
\begin{equation}
p(x)=x^2-x +{\gamma^2\over 2}= (x-\alpha^2_+)(x-\alpha_-^2)
\end{equation}
are given by 
\begin{equation}
\alpha^2_\pm ={1\over 2}(1\pm\sqrt{1- 2\gamma^2})
\end{equation}
%
Then the above equation can be solved by the integral
\begin{equation}
\mu_0\pm \mu=\int^\infty_{\sqrt{f}} {dx\over
\sqrt{(x^2-\alpha^2_+)(x^2-\alpha_-^2)}}={1\over \alpha_+}
\int_0^{{\alpha_+\over \sqrt{f}}} {dx\over
\sqrt{(1-x^2)(1-k^2 x^2)}}
\end{equation}
where $k=\alpha_-/\alpha_+$ and $\alpha_+\mu_0 = K(k)$.
We choose here $\mu_0$ such that $\mu=0$ at the turning
point. 
Then the coordinate $\mu$ is ranged over the interval
$[-\mu_0, \mu_0]$, where one can show that
$\mu_0  \ge \pi/2$
for any $\gamma$ in $[-{1\over \sqrt{2}} ,{1\over \sqrt{2}}]$.
With the help of the elliptic
integral of the first kind,
\begin{equation}
F(\varphi, k)=
\int^\varphi_0 {d\alpha\over
\sqrt{1-k^2 \sin^2\alpha}} \,,
\end{equation}
the  above integral may be represented by
\begin{equation}
\mu_0\pm \mu={1\over \alpha_+}
F\left(\sin^{-1} \Bigl({\alpha_+\over f^{1\over 2}}\Bigr), k\right)\,.
\end{equation}
One may invert the above expression
as \cite{Freedman}
\begin{equation}
f= {\alpha_+^2\over {\rm sn}^2 (\alpha_+ (\mu+\mu_0),k)}
\end{equation}
using the Jacobi elliptic functions,
${\rm sn}(z, k)$,  defined by
\begin{equation}
z=\int_0^{{\rm sn}(z,k)} {dx\over
\sqrt{(1-x^2)(1-k^2 x^2)}} \,.
\end{equation}
The cosine amplitude ${\rm cn}(x,k)$ and the delta amplitude
can be introduced by the relations,
\begin{equation}
{\rm cn}(z,k)= \cos (\sin^{-1} ({\rm sn}(z,k)))\,,\ \ \ \ \
{\rm dn}(z,k)= \sqrt{1-k^2 {\rm sn}^2(z,k)}\,.
\end{equation}
Then
the dilaton can be integrated explicitly as
\begin{equation}
\phi=\phi_0+ \sqrt{2}\ln \Bigl( {\rm dn}(\alpha_+ (\mu+\mu_0),k)
-k\,\,{\rm cn}(\alpha_+ (\mu+\mu_0),k)
\Bigr)\,.
\end{equation}

In fact using a different coordinate defined by
\begin{equation}
y = \int^\mu_0 ds \sqrt{f(s)}\,,
\end{equation}
the solution may be presented in terms of
elementary functions. In this coordinate, the
three dimensional metric $g_{ab}$ in (\ref{reduction}) takes the form
\begin{equation}
ds_3^2 = f(y) ds^2_{AdS_2} + dy^2\,.
\end{equation}

It is straightforward to find the solution\cite{Freedman}
\begin{eqnarray}
&& f(y)= {1\over 2} (1+ \sqrt{1-2 \gamma^2} \cosh 2y)\,,\nn\\
&& \phi = \phi_0 +{1\over \sqrt{2}}
\ln\left({1+\sqrt{1-2\gamma^2}+ \sqrt{2}\gamma\tanh y\over
1+\sqrt{1-2\gamma^2}-\sqrt{2}\gamma\tanh y}\right)\,.
\label{freed}
\end{eqnarray}
Note that the boundary values of the dilaton at $\mu=\pm \mu_0$
are evaluated   as
\begin{eqnarray}
\phi_\pm-\phi_0 =\pm {1\over \sqrt{2}} \tanh^{-1}  \sqrt{2}\gamma
=\pm {1\over 2\sqrt{2}}\ln \left({1+\sqrt{2}\gamma\over
1-\sqrt{2}\gamma
}\right)
\,. \label{phijump}
\end{eqnarray}

The IIB string theory has the $SL(2,Z)$ duality
symmetry and the classical IIB
supergravity possesses the $SL(2,R)$ symmetry. Hence by performing the
$SL(2,R)$ transformation, one may generate the new family of
solutions. These solutions in general involve the
nonvanishing axion $\chi$ and NS-NS three form field
strength in addition. Here we shall not present the explicit form of such
solutions generated by the $SL(2,R)$ transformation, but would like to note that
the corresponding  dual CFT involves axionic domain wall together with
the  jump of the coupling. Namely in the dual CFT,
the $\theta$ angle  jumps  too at the interface.

For later comparison let us compute the one-point function of dual
dilaton operators.
By introducing  the Poincare patch parametrization
for the AdS$_2$ part,
the three
metric  reads
\begin{eqnarray}
ds_3^2=
{f(\mu)\over y^2}
 \left( y^2d\mu^2  -dx_0^2+ dy^2\right)
\,.
\end{eqnarray}
We adopt the conformal compactification where the scaling factor
is given by $\sqrt{f}/y$.
Combining two halves of $R^2$ defined by $\mu=\pm \mu_0$,
the boundary becomes a full
$R^2$,
on which the dual Janus CFT is defined. In the near boundary region,
the above metric can be rewritten
as
\begin{eqnarray}
ds_3^2=
{1\over z^2}
 \left(dz^2  -dx_0^2+ dx^2
+ O({z^2\over x^2} dx_a dx_b)
\right)
\,,
\end{eqnarray}
where $z= y\, {\rm sn} (\alpha_+(\mu_0-\mu), k)/\alpha_+$
is the inverse
of the scale factor
and $x=y\,{\rm cn} (\alpha_+(\mu_0-\mu), k)$.
Note that  the dilaton behaves near the boundary  as
\begin{eqnarray}
\phi=\phi_\pm \mp{\gamma\over 2}(\mu\mp\mu_0)^2 + \cdots
=\phi_\pm -{\gamma\over 2}{\epsilon(x) z^2/x^2} + \cdots
\,.
\end{eqnarray}

Using the AdS/CFT correspondence, we have the relation
\begin{eqnarray}
\langle O_\phi\rangle = {\delta I_\phi\over \delta \phi}
= -{1\over 16
\pi G_3} \sqrt{g_3} g_3^{zz}\partial_z \phi |_{z=0}
= - {\gamma\, Q_1 Q_5\over 4\pi} {\epsilon(x)\over x^2}\,,
\end{eqnarray}
where we take the dual operator as
\begin{eqnarray}
O_{\phi}(z,\bar z) = -{1\over 4\pi } \sum_{i,a}:\partial X_{i,a} \bar
\partial X_{i,a}:\,.
\end{eqnarray}

In Ref.~\cite{Papa}, the Fefferman-Graham coordinate system for
the above metric
is obtained and the boundary perturbation of the metric can be identified.
Using this, one can show that
\begin{eqnarray}
\langle T_{ab}\rangle = 0\,,
\end{eqnarray}
which is an expected result.


\section{The dual CFT}
In this section we review the two dimensional CFT dual
of type IIB string theory on $AdS_{3}\times S^{3}\times M_{4}$
 where $M_{4}$ is either $T_{4}$ or $K_{3}$.  The
central charge of the CFT can be related to the $AdS_{3}$
curvature by
\begin{equation}\label{cchargea}
c={3R_{ads}\over 2 G_3}\,.
\end{equation}

This background can be obtained from a near horizon limit
of $Q_{1}$ D1-branes and $Q_{5}$ D5-branes, wrapping $M_{4}$.
The  theory living on the D1-D5 common $1+1$ dimensional
worldvolume is a $N=(4,4)$  supersymmetric field theory.
The CFT dual to the near horizon limit of the D1-D5 system
\cite{Maldacena:1997re} is the IR fixed point of the $N=(4,4)$ theory.
This theory can be described as a $1+1$ dimensional
supersymmetric $\sigma$-model where the target space is
the moduli space of $Q_{1}$ instantons in a two dimensional
 $SU(Q_{5})$ gauge theory.  The moduli space is $4n$
dimensional where $n= Q_{1}Q_{5}$ (for $M_{4}=T_{4}$) or
$n=Q_{1 }Q_{5}+1$ (for $M_{4}=K_{3}$). The conformal field
 theory is given
by \cite{Vafa:1995bm}\cite{Witten:1997yu}\cite{Dijkgraaf:1998gf}
 the smooth resolution  of the orbifold CFT of the
symmetric product  $M^{n}/S_{n}$. In the following we focus
 on the case where $M_{4}= T^{4}$.  The central charge
(\ref{cchargea}) of the CFT is then $c= 6n = 6 Q_{1}Q_{5}$.
The orbifold $T_{4}^{n}/S_{n}$ can be constructed by
starting with the free field CFT representing the tensor
product $T_{4}^{n}$
\begin{equation}\label{orbcft}
S= {1\over 2 \pi \alpha'}\int d^{2}z \sum_{i,a}
\Big( \partial X_{i,a} \bar \partial X_{i,a} + \psi_{i,a }\bar
\partial \psi_{i,a}+ \bar \psi_{i,a }\partial \bar \psi_{i,a}\Big)\ .
\end{equation}
The indices $i=1,2,\cdots 4$, and $a=1,2,\cdots n$
parameterize  $n$ copies of the four torus $T^{4}$.
Hereafter we shall set $\alpha'=2$.
The orbifold then projects onto states which are invariant under
$S_{n}$ acting by permutation on the coordinates $X_{i,a}$.
Modular invariance mandates the inclusion of twisted sectors
 which contain marginal operators responsible for the smooth
resolution of the orbifold singularities.
The correlators of the unperturbed orbifold CFT for the
 bosonic fields is given by
\begin{equation}
\langle X^{i,a}(z,\bar z)X^{j,b}(w,\bar w)\rangle= -
\ln \mid z-w\mid^{2}   \delta^{{ij}} \delta^{a\, b}\ ,
\end{equation}
where the sum is over permutations of the index $b$
following the standard orbifolding procedure.

The Janus deformation of $AdS_{3}\times S_{3}\times M_{4}$
has a nontrivial profile for the six dimensional dilaton
$\phi_{(6)}$. In order to identify the dual of the Janus
solution, one first has to identify the operator dual to
the dilaton. Symmetry considerations simplify the identification,
 the dilaton  in the Janus solution does not depend on the
 coordinates of the $S_{3}$ or $T_{4}$, in the dual CFT
this means that the operator transforms trivially under
the $SU(2)\times SU(2)$ R-symmetry and $SU(2)\times SU(2)$
global symmetry of the $N=(4,4)$ SCFT. Furthermore the
constant Kaluza-Klein mode on the sphere of the dilaton
is a massless scalar field in $AdS_{3}$ and hence
corresponds to a marginal deformation with conformal
dimensions $(h,\bar h)=(1,1)$. A natural guess for the
dual operator is therefore:
\begin{equation}\label{opdefa}
O_{\phi}(z,\bar z) = -{1\over 4\pi}
\sum_{i,a}:\partial X_{i,a} \bar \partial X_{i,a}:
\end{equation}
That the operator 
has the correct conformal dimensions can be seen from the
two point function
\begin{equation}
\langle O_{\phi} (z,\bar z) O_{\phi} (w,\bar w)\rangle =
 {n\over 4\pi^2 \mid z-w\mid ^{4}}
\end{equation}

As discussed in the previous section the solution (\ref{ansatza})
incorporates  a Janus type $AdS_{2}$ slicing of $AdS_{3}$.
 The holographic dual theory is therefore an interface CFT with
two half-spaces glued together by a one dimensional interface.
 Furthermore the dilaton takes two values  $\phi_{\pm}$
(\ref{phijump}) at the boundary $\mu=\pm \mu_{0}$ corresponding
to the two half-spaces.

The  appearance of the dilaton factor in front of the
$AdS_{3}$ part of the metric (\ref{ansatza}) might worry
the reader since this implies that asymptotically  the
 $AdS_{3}$ curvature radius behaves as
 $R_{ads}\sim  e^{{1\over 2} \phi_{\pm}}$ near the two boundary
components. Does this imply that the central charge
(\ref{cchargea}) is jumping across the defect? This would
 clearly be strange since the Janus deformation is associated
with a marginal operator which should not change the central
charge of the CFT. The resolution of this puzzle lies in the
fact that the three dimensional Newton's constant is behaving
like $G_3  \sim e^{{1\over 2} \phi_{\pm}}$ near the
boundary and the dilaton factors cancel out in the formula
for the central charge.

The  $AdS_{3}$ Janus deformation  can be analyzed using
conformal perturbation theory. For the location of the
interface at $x_{2}=0$, where $z=x_{1}+i x_{2}$, the
deformation is defined by adding the following term
to the action
\begin{equation}
S=S+ \lambda \int d^{2}z \epsilon(x_{2}) O_{\phi} (z,\bar z)
\end{equation}
where $\lambda= \gamma + O(\gamma^2)$.
We can apply conformal perturbation theory method which was
applied for the $AdS_{5}$ Janus solution in  \cite{Karch}.
 We will only calculate the simplest correlation functions
 which provide  nontrivial checks of the
correspondence\footnote{In Ref.~\cite{Osborn},
some of the correlation functions are computed
exactly  for the ICFT.}.
First,
it is clear that the expectation value $\langle O_{\phi}(z,\bar z)\rangle =0$
of the operator (\ref{opdefa}) vanishes in the unperturbed theory
since the operator $O_{\phi}$ is normal ordered.
Second the one point function of  (\ref{opdefa})  to order
$o(\lambda)$ is given by
\begin{eqnarray}
\langle O_{\phi} (w,\bar w) \rangle_{\lambda}&=& \lambda
\langle O_{\phi} (w,\bar w) \int d^{2}z \epsilon(x_{2})
 O_{\phi} (z,\bar z) \rangle +o(\lambda^{2}) \nonumber \\
 &=&  -{\gamma n\over 4\pi} {\epsilon(w_{2}) \over \mid w\mid ^{2}}
\end{eqnarray}


Third the expectation value of the energy momentum tensor to
first order in $\lambda$ is given by
\begin{eqnarray}
\langle T(w) \rangle_{\lambda}&=&  \lambda \langle T(w)
\int d^{2}z \;\epsilon(x_{2})   O_{\phi} (z,\bar z) \rangle
 =0  \nonumber \\
\langle \bar T(\bar w) \rangle_{\lambda}&=&  \lambda \langle
\bar T({\bar w}) \int d^{2}z \;\epsilon(x_{2})   O_{\phi} (z,\bar z) \rangle
=  0
\end{eqnarray}


\section{Time dependent BTZ-type black hole}

In this section, we would like to present another type
of related
gravity solution. This solution describes a black hole of a BTZ
type. But the black hole solution has an unconventional character.
The horizon size and the dilaton value on the horizon are time
dependent. The geometry involves two disconnected boundaries
and the couplings of the boundaries differ from each other.
A similar kind of  multi boundary solution
in the Euclidean context
is mentioned in Ref.~\cite{Maoz} but the geometry there
is not directly related the one presented here.

The construction of the solution goes as follows.
Note that the ansatz
in (\ref{ansatza}) may be equivalently presented as
 \begin{eqnarray}
ds_3^2  = f(\mu)\cos^2\mu\, ds^2_{AdS_3}
\ ,\ \
\phi=\phi(\mu)\ , \label{ansatzb}
\end{eqnarray}
with the $AdS_3$ metric
\begin{eqnarray}
ds^2_{AdS_3}= {1\over \cos^2\mu}(d\mu^2 + ds^2_{AdS_2})\,.
\end{eqnarray}
Then, of course, the dilaton and the conformal factor are solved by the
Janus solution of the previous section.
As in the $AdS_5$ case, the $AdS_3$ becomes the
global/Poincare metric if one uses the global/Poincare parametrization
for the $AdS_2$. In fact one may replace $ds^2_{AdS_3}$ by any
three metric satisfying $R_{ab}= -2 g_{ab}$,
where a translation in $\mu$ should be isometry of the metric
$\cos^2\mu\, g_{ab}$.
 The $AdS_3$ metric
for instance can be replaced by the metric for any BTZ black hole.
Here we illustrate the detailed  construction for the
zero angular momentum case only. The BTZ black hole solution
may be constructed using the orbifolding technique. Note that
$AdS_3$ space is the hyperboloid in $R^{2,2}$ satisfying
\begin{eqnarray}
-Y_0^2-Y_3^2+ Y_1^2+Y_2^2=-1\,.
\end{eqnarray}
We then use the parametrization of the $AdS_3$ space by
\begin{eqnarray}
&& Y_0= \sqrt{{r^2\over r^2_0}-1}\, \sinh r_0 t\,,  \ Y_2= \pm
\sqrt{(r/r_0)^2-1}\, \cosh r_0 t\  \nn \\
&& Y_1= {r\over r_0} \sinh r_0 \theta\,, \ \ \ \ \ \ \ \ Y_3= \pm
{r\over r_0} \cosh r_0 \theta\,.
\end{eqnarray}
The metric takes the form of
\begin{eqnarray}
ds^2_{BTZ}= -(r^2-r_0^2) dt^2 + {dr^2\over r^2-r_0^2} + r^2 d\theta^2\,.
\end{eqnarray}
With the identification  $\theta \sim \theta + 2\pi$, the above
 describes the BTZ black hole with vanishing angular
momentum\cite{BTZ}.  The horizon is at $r=r_0$
and $r=0$ corresponds to
a singularity of the orbifold type.
In the solution, the coordinate $\mu$ is related to the BTZ coordinates
by
\begin{eqnarray}
\tan \mu = Y_2=
\pm
\sqrt{(r/r_0)^2-1}\, \cosh r_0 t\
\,.
\end{eqnarray}
The BTZ coordinate does not cover the whole region of our geometry; it
can be also extended to the asymptotic region in addition
 to the region beyond the horizon. To see this, let us introduce
the Kruskal
coordinates
\begin{eqnarray}
V= e^{r_0(t+ r_*)}\,, \ \ \ \ \  U=- e^{-r_0(t- r_*)}
\,,
\end{eqnarray}
where $r_*$ denotes
\begin{eqnarray}
r_* = {1\over 2r_0} \ln \left({r-r_0\over r+ r_0}\right)
\,.
\end{eqnarray}
The coordinates $r$ and $\mu$ are related to $(U, V)$ by
\begin{eqnarray}
{r \over r_0} = {1-UV\over 1+ UV}\,, \ \ \ \ \cos^2\mu =
{(1+UV)^2\over (1+U^2)(1+V^2)}
\,.
\end{eqnarray}
In this coordinate,
the three metric becomes
\begin{eqnarray}
ds_3^2= {f(\mu) \over (1+ U^2)(1+V^2)}(-4dUdV + r_0^2 (1-UV)^2 d\theta^2)
\,.
\end{eqnarray}
with $U,\, V  \in (-\infty, \infty)$ as a result of the extension.
But even this new coordinate does not cover the whole geometry and can be
extended further to the asymptotic region. The fully extended
geometry may be obtained by introducing a parametrization,
\begin{eqnarray}
V=\tan w_1\,, \ \ \ \ \ \  U= \tan w_2\
\,.
\end{eqnarray}
The metric now takes the form
\begin{eqnarray}
ds_3^2= f(\mu)(-4dw_1dw_2 + r_0^2 \cos^2(w_1+w_2)
d\theta^2)=  f(\mu)(-d\tau^2 +d\mu^2 + r_0^2\cos^2\tau
d\theta^2)
\,,
\label{tbh}
\end{eqnarray}
where $\tau= w_1+w_2$ and $\mu=w_1-w_2$. One may use
$y=\int^\mu_0 ds\sqrt{f(s)}$ that is introduced before and the metric
is then represented in terms of elementary functions  by
\begin{eqnarray}
ds_3^2=  dy^2+ {1\over 2} \left(1+\sqrt{1-2\gamma^2} \cosh 2y\right)
(-d\tau^2 + r_0^2\,\cos^2\tau\,
d\theta^2)
\,,
\end{eqnarray}
with the scalar field given in (\ref{freed}).
The orbifold singularity is now at $\tau=\pm \pi/2$ and the asymptotic
spatial infinities are located
at $\mu=\pm \mu_0$. Thus the Penrose diagram is covering the
region $\tau \in [-\pi/2, \pi/2]$ and $\mu\in [-\mu_0, \mu_0]$.

\begin{figure}[ht!]
\centering \epsfysize=9cm
\includegraphics[scale=0.8]{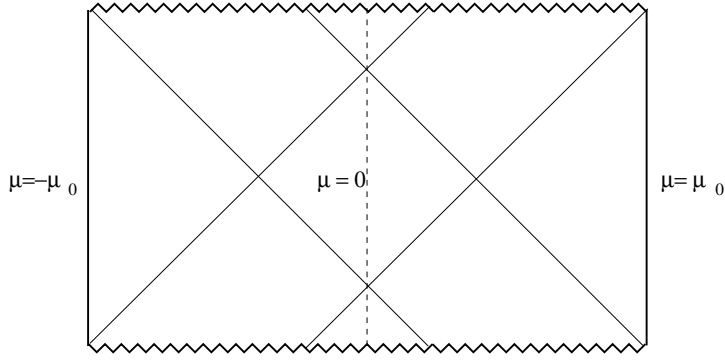}
\caption{\small Penrose diagram for the time dependent  black
hole with two
couplings. The $\tau$ ($ \in [-\pi/2,\pi/2]$) coordinate runs
vertically upward and $\mu$ ($ \in [-\mu_0,\mu_0]$) to the right
horizontally}
\end{figure}

The coupling on the right/left side boundary takes the value of
$e^{\phi_+}$/$\,e^{\phi_-}$. Two different boundary CFT's are
correlated  through
the bulk in a subtle manner\cite{Maoz}.
The boundary  CFT is not the Janus type. Rather the coupling is
uniform on the entire circle and remains constant in time.
The geometry has the time reversal
symmetry at $\tau=0$ axis and a parity symmetry
under the interchange of $\mu$ to $-\mu$.  In the right asymptotic regions
with $\tau \ge 0$, the future-horizon area is described by the points
$\mu=\mu_0-\pi/2 +\tau$ with $\tau\ \in\ [0, \pi/2]$. The horizon
area may be evaluated  as
\begin{eqnarray}
A(\tau)= 2\pi r_0 {\alpha_+ \sin(\pi/2-\tau)\over
{\rm sn}(\alpha_+(\pi/2-\tau),k)}
\,.
\end{eqnarray}
For $\tau \in [0, \pi/2]$, the area grows monotonically in time and
the minimum value is $2\pi r_0 {\alpha_+ \over
{\rm sn}\left({\pi \alpha_+\over 2},k\right)}$ and the maximum,
$2\pi r_0$.
%

Finally the case of $\gamma^2 \ge 1/2 $ shall be treated 
when we consider the black holes in general dimensions.

\section{Time dependent topological black holes in higher
dimensions}
In this section, we would like to generalize the black
 hole solution of previous section to  higher
dimensions. The construction is again fairly straightforward.
We begin with a dilaton Einstein gravity described by
\begin{equation}
I=  {1\over 16\pi G_d}\int \sqrt{g_d}\left(R_d-
g_d^{ab}\partial_a\phi
\partial_b\phi + (d-1)(d-2)
\right)\ .
\label{daction}
\end{equation}
with $d\ge 3$.
As we just described in the previous sections, any
solutions of the above action
for  the $d=3$ case can be embedded
into the 10d type IIB supergravity. This is also true for
the $d=5$ case, which leads to a deformation of
$AdS_5\times S^5$ geometry.
The ansatz  may be  taken as
 \begin{eqnarray}
ds_d^2  = f(\mu)(d\mu^2 +ds^2_{d-1})
\ ,\ \
\phi=\phi(\mu)\ , \label{ansatzc}
\end{eqnarray}
where the $(d-1)$ dimensional metric $\tilde{g}_{ij}$
describes an Einstein space satisfying
$\tilde{R}_{ij}= -(d-2)\tilde{g}_{ij}$.
The equation of motion for the dilaton can be integrated
leading to
\begin{equation}
\phi'(\mu) = {\gamma\over f^{d-2\over 2}(\mu)}\,,
\label{dilac}
\end{equation}
and the Einstein equations are reduced to
\begin{equation}
f'f' = 4 f^3 -4 f^2+{4\gamma^2\over (d-1)(d-2)} f^{4-d}\ .
\label{einsteinc}
\end{equation}
This can be solved by the integral
\begin{equation}
\mu_0\pm \mu=\int^\infty_{f} {dx\over 2
\sqrt{x^3-x^2+{\gamma^2\over (d-1)(d-2)} x^{4-d}}}\,,
\end{equation}
where $\mu_0$ is chosen such that $\mu=0$ at the turning
point. 
Here we are discussing the  case of
$\gamma^2 \le \gamma_c^2 $ with $\gamma_c^2=
(d-2)\left({d-2\over d-1}\right)^{d-2}$, for which
the geometry is free of timelike curvature singularity. 
Then the coordinate $\mu$ is  ranged over the interval
$[-\mu_0, \mu_0]$ with
$\mu_0  \ge \pi/2$.

Up to this point, there is no difference from the
construction of the Janus solutions except that
we put the $(d-1)$ dimensional spacetime
in a generic form of the Einstein manifold with negative
cosmological constant.
Now the trick is to take $\tilde{g}_{ij}$ as
\begin{equation}
ds^2_{d-1}=
-d\tau^2 + \cos^2\tau ds^2_\Sigma
\end{equation}
where $ds^2_\Sigma$ is describing the compact, smooth,
finite volume Einstein space metric in $(d-2)$
dimensions satisfying
$R^\Sigma_{kl}= -(d-3) g^{\Sigma}_{kl}$.
One example of such space is given by the quotient of
the hyperbolic space $H_{d-2}$ by a discrete subgroup of
the hyperbolic symmetry group, $SO(1,d-2)$. One can pick
the subgroup $\Gamma$ such that $\Sigma_{d-2}=H_{d-2}/\Gamma$
is a compact, smooth, finite volume space. Notice that
$\Sigma_2$ constructed
this way corresponds to constant curvature Riemann surface of
genus no less than two. 

The resulting metric,
\begin{eqnarray}
ds_d^2=  f(\mu)(-d\tau^2 +d\mu^2 + \cos^2\tau
ds_\Sigma^2)
\,,
\label{tbhc}
\end{eqnarray}
is the d dimensional  generalization
of the three metric in (\ref{tbh}).
 Note here that
$\tau\in [-\pi/2, \pi/2]$ as before. Therefore
the Penrose digram for this higher dimensional
black hole is again described by Figure~1, in which
a point represents $\Sigma$ slice.
The spacetime is locally isomorphic to $AdS_d$ and
the curvature singularity at $\tau=0$ is again
of the orbifold type.

Let us  turn to the over-critical case of $\gamma^2 > \gamma_c^2$.
The scale factor $f(\mu)$ is now ranged over $[0,\infty)$ without
any  turning point and the 
geometry involves an extra curvature singularity 
at $f=0$, which is timelike. The factor $f(\mu)$ can be 
solved by the integral
\begin{equation}
\mu_0- \mu=\int^\infty_{f} {dx\over 2
\sqrt{x^3-x^2+{\gamma^2\over (d-1)(d-2)} x^{4-d}}}\,.
\end{equation}
where $\mu_0$ can be taken to be arbitrary.
Since $f$ has to be non negative, the $\mu$ coordinate
is ranged over $[\mu_0-\kappa,\mu_0]$ where the length of the 
interval, $\kappa(\gamma^2)$, is determined  by the integral
\begin{equation}
\kappa(\gamma^2)=\int^\infty_{0} {dx\over 2
\sqrt{x^3-x^2+{\gamma^2\over (d-1)(d-2)} x^{4-d}}}\,.
\end{equation}
For $\gamma^2 > \gamma_c^2$, $\kappa(\gamma^2)$
is decreasing monotonically from infinity to zero.
The metric for the geometry  is still given by (\ref{tbhc})
but the timelike singularity occurs at $\mu=\mu_0 -\kappa$.
Because of this, the spacetime cannot be extended to the region
of $\mu < \mu_0 -\kappa$.  If $\gamma_c^2 < \gamma^2 < \gamma_s^2$
with $\gamma_s^2$ defined by $\pi/2=\kappa(\gamma_s^2)$,
the geometry becomes free of naked singularity describing a 
regular time-dependent black hole.

Representing the volume of $\Sigma$ space by ${\cal V}_\Sigma$,
the future-horizon area is given by
\begin{eqnarray}
A(\tau)= {\cal V}_\Sigma
(\cos(\tau) f^{1\over 2}(\mu_0 +\tau-\pi/2))^{d-2}
\,.
\end{eqnarray}
One can check that the area is monotonically
increasing for $\tau \in [-\pi/2, \pi/2]$ reaching
the maximal value ${\cal V}_\Sigma$ at
$\tau=\pi/2$. In this sense, the black hole is truly
time dependent for nonvanishing $\gamma$.
For $\gamma=0$, the horizon area remains constant and the black
hole
becomes static. In fact, it is straightforward to show that
the $\gamma=0$ solution corresponds to the
$M=0$ and $k=-1$ topological black hole solution of
Ref.~\cite{Birmingham}.

We shall not discuss the detailed framework for the
gravity/gauge theory correspondence here. Note, however,
that the boundary metric of the dual CFT can be obtained
by the multiplication of
any $h^2$ where
$h$ approaches  linearly zero at the boundary. Different choice of
$h$ leads to a different boundary metric.
By taking $h= f^{-{1\over 2}}$ for instance,
the boundary metric becomes
\begin{eqnarray}
ds_B^2=  -d\tau^2+\cos^2\tau
ds_\Sigma^2
\,,
\label{bdmetric}
\end{eqnarray}
which is cosmological.
The dual CFT will be defined in this cosmological
background spacetime. Although the boundary spacetime reveals
the big-bang and big-crunch singularities at $\tau=\pm\pi/2$,
the bulk-extended metric at the points is perfectly
regular.

If one chooses $h = f^{-{1\over 2}}/\cos\tau$, the boundary
now becomes
\begin{eqnarray}
ds_B^2=  -dt^2+
ds_\Sigma^2
\,,
\label{bdmetric1}
\end{eqnarray}
where $t\in (-\infty,\infty)$. Thus the dual CFT
is defined on $R\times \Sigma$ now.\footnote{The $\gamma=0$ case of
  the correspondence is discussed in the Refs. \cite{Emparan:1999gf,
    Buchel:2004rr}.  }
In this case, the finite temperature system starts off with 
some out of
equilibrium state at $\tau=0$ and then the excess kinetic energy
is thermalized reaching the equilibrium at late time.
One  expects that detailed information about
the thermalization process can be extracted from the behavior of the
solution.
Further study is necessary in this direction.

\section{Conclusions}

We generalized our previous study of the Janus deformation
to the $AdS_3\times S^3\times M_4$ space. The $AdS_3$ part
is replaced by the $AdS_2$-sliced Janus.
However, the total spacetime is not the simple product of
$\mbox{Janus}_3\times S^3\times M_4$, as one might have thought.
 Indeed each component of the product space is warped in a
specific manner by an exponential of the dilaton. Besides
this nontriviality, the solution is expressed in a simple
analytic form. Thus one may hope that further quantitative
studies will be much facilitated.
The dual CFT interpretation is similar to the $AdS_5\times S^5$
case. A spatially dependent marginal deformation dual to
the dilaton leaves the conformal invariance only in the
interface of two halves of the boundary space.  The resulting
dual field theory is an ICFT$_2$.

Apart from the Janus deformation of the $AdS_3\times S^3\times M_4$
space, we also discussed the Janus deformation of the BTZ
black hole. The Janus BTZ black hole turned out to be time
dependent and has two disconnected boundaries. The dilaton
does not divide each boundary component into two halves.
Rather, it takes one value in one component of the
boundaries and the other in the other component.
This black hole solutions can be generalized to the higher dimensions.
Among these the three and five dimensional ones can be embedded
into the type IIB supergravity.
It would
 be quite interesting to study further the microscopic
description of these time dependent  black holes.

\section*{Acknowledgements}
We are grateful to Andreas Karch and Soo-Jong Rey for
useful discussions and
conversations.
DB would like to thank the Particle Theory Group of
University of Washington for the warm  hospitality.
The work of DB is supported in part by
 KOSEF ABRL R14-2003-012-01002-0 and KOSEF SRC CQUeST R11-2005-021. The work of MG is supported in part by NSF grant PHY-04-56200.

\end{document}